\documentclass[twoside,epsf]{article}
\usepackage{graphicx}
\newcommand{\lascia}[1]{}

\newcount\Mac  \Mac=0  
\newcommand{\ifMac}[2]{\ifnum\Mac=1 #1 \else #2 \fi}

\newcommand{\eV}{\,{\rm eV}}
\newcommand{\cm}{\,{\rm cm}}

\def\Red  {}

\def\Black{}
\def\Blue {}

\newcommand{\Jhep}{{\rm J.hep}}

\newcommand{\fig}[1]{~\ref{fig:#1}}

\newcommand{\GeV}{\,{\rm GeV}}

\newcommand{\MeV}{\,{\rm MeV}}

\newcommand{\One}{\hbox{1\kern-.24em I}}
\newcommand{\NP}{Nucl. Phys.}
\newcommand{\PRL}{Phys. Rev. Lett.}
\newcommand{\PL}{Phys. Lett.}
\newcommand{\PR}{Phys. Rev.}

\newcommand{\eq}[1]{~(\ref{eq:#1})}
\def\diag{\mathop{\rm diag}}

\def\circa#1{\,\raise.3ex\hbox{$#1$\kern-.75em\lower1ex\hbox{$\sim$}}\,}
\makeatletter
\def\art{\@ifnextchar[{\eart}{\oart}}
\def\eart[#1]#2#3#4#5#6{{\rm #2}, {\em #3 \bf #4} {\rm (#6) #5} ({\em #1})}
\def\hepart[#1]#2{{\rm #2, \em#1}}
\newcommand{\oart}[5]{{\rm #1}, {\em #2 \bf #3} {\rm (#5) #4}}

\newcounter{alphaequation}[equation]
\def\thealphaequation{\theequation\hbox to
0.6em{\hfil\alph{alphaequation}\hfil}}
\def\eqnsystem#1{
\def\@eqnnum{{\rm (\thealphaequation)}}
\def\@@eqncr{\let\@tempa\relax \ifcase\@eqcnt \def\@tempa{& & &} \or
  \def\@tempa{& &}\or \def\@tempa{&}\fi\@tempa
  \if@eqnsw\@eqnnum\refstepcounter{alphaequation}\fi
\global\@eqnswtrue\global\@eqcnt=0\cr}
\refstepcounter{equation} \let\@currentlabel\theequation \def\@tempb{#1}
\ifx\@tempb\empty\else\label{#1}\fi
\refstepcounter{alphaequation}
\let\@currentlabel\thealphaequation
\global\@eqnswtrue\global\@eqcnt=0 \tabskip\@centering\let\\=\@eqncr
$$\halign to \displaywidth\bgroup \@eqnsel\hskip\@centering
$\displaystyle\tabskip\z@{##}$&\global\@eqcnt\@ne
\hskip2\arraycolsep\hfil${##}$\hfil& \global\@eqcnt\tw@\hskip2\arraycolsep
$\displaystyle\tabskip\z@{##}$\hfil
\tabskip\@centering&\llap{##}\tabskip\z@\cr}
\def\endeqnsystem{\@@eqncr\egroup$$\global\@ignoretrue} \makeatother

\newcount\n

\begin{document}\twocolumn[
\centerline{hep-ph/0011307 \hfill CERN--TH/2000--345 \hfill IFUP--TH/2000--43 \hfill SNS-PH/00--17}
\vspace{5mm}

\Black
\vspace{0.5cm}
\centerline{\LARGE\bf\Red Non standard analysis of the solar neutrino
anomaly}\centerline{\large\bf(updated on July 2001 including the SNO CC data$^*$)}
\centerline{\large\bf(updated on July 2002 including the SNO day/night spectral CC and NC data$^\dagger$)}
\Black

\medskip\bigskip\Black
\centerline{\large\bf Riccardo Barbieri}\vspace{0.2cm}
\centerline{\em Scuola Normale Superiore, Piazza dei Cavalieri 7, I-56126
Pisa, Italy and INFN}
\vspace{3mm}
\centerline{\large\bf Alessandro Strumia}\vspace{0.2cm}
\centerline{\em Theory division, CERN and Dipartimento di Fisica, Universit\`a di Pisa and INFN}
\vspace{1cm}
\Blue\centerline{\large\bf Abstract}
\begin{quote}\large\indent
Continuing previous work, a model independent analysis of the solar
neutrino anomaly
is performed in terms of neutrino oscillations, allowing a comparison with
the predictions of the
Standard Solar Model. SMA and LMA solutions emerge also in this case,
although somewhat different from
the standard ones. The significance of the NC/CC double ratio measurable in
SNO is illustrated in this context.

\end{quote}\Black
\vspace{0.5cm}]

\footnotetext[1]{The addendum at page \pageref{6in} (section 5)
is not present in the published version of this paper.}
\footnotetext[2]{The addendum at page \pageref{7in} (section 6)
is not present in the published version of this paper.}

\section{Introduction and motivations}
Flavour neutrino oscillations continue to be a pretty controversial matter.
It is fair to recall that,
so far, no direct signal of them, like neutrino appearance or explicit
oscillation patterns, have  been observed
either in solar or in atmospheric neutrinos.
The up/down asymmetry of the flux of the atmospheric $\nu_\mu$ and
$\bar{\nu}_\mu$ neutrinos
gives, however,
an indisputable evidence for the presence of an atmospheric neutrino anomaly.
No equally clear evidence has been found, on the other hand, for the solar
neutrino anomaly~\cite{ClSun,KaSun,GaSun,ExpsSun}, since
its standard interpretation  relies on a combination of
many different experimental and theoretical ingredients.
Furthermore, the LNSD result still awaits for an independent confirmation.

On the solar neutrinos, which are the subject of this paper, quite
different attitudes
can be taken, depending on the weight one gives to the input of the
Standard Solar Model (SSM)~\cite{BP98}.
On one side,
it does not look reasonable to consider the solar neutrino anomaly as an
artifact due
to a large unknown error
in solar models or in solar neutrino experiments.
The other extreme attitude is to assume that all the ingredients of the
analysis 
are correct, thus obtaining a rather precise determination of the neutrino
oscillation parameters.
As well known, the best fits of the solar neutrino deficit in this
framework are given by few peculiar
energy-dependent survival probabilities.

The truth is that unfortunately, so far, SuperKamio\-kande (SK) has not
found any evidence for a distortion of the energy spectrum,
nor for Earth regeneration effects, nor for seasonal variations of the
neutrino flux.
Furthermore, the most recent SK data~\cite{KaSun,ExpsSun} worsen the
quality of the best fit,
with the net result
that the new best fit regions now include values
of the oscillation parameters previously discarded on the basis of the sole
neutrino rates.
Recent analyses found that
all the distinct best fit solutions have a high
goodness-of-fit probability~\cite{recentfits}. However, at least in part,
 this is just a reflection of having fitted the few really problematic
data together with many other `degrees of
freedom' that have not much to do with the problem.
To really judge the quality of the fit one should
perform a more complete statistical analysis or
rewrite the data in terms of a minimal set of `optimal'
observables\footnote{A similar comment can
be done for atmospheric neutrinos.
It is hard to judge if the $\nu_\tau \to \nu_{\rm sterile}$ interpretation
gives an acceptable fit
by looking only at the minimal $\chi^2$ of a global fit that includes
electron data,
low energy data and too many zenith angle bins.}.

%

In view of this situation,
we find it useful to come back to an analysis which has minimal dependence
upon the
SSM inputs. This is the purpose of this paper, continuing previous work
along similar lines.
From an experimental point of view, the main new information comes from the
SK measurements, mentioned above, of the energy spectrum and
of day/night or seasonal variations of the neutrino flux.
Their interpretation has little to do with the theoretical input of the SSM.

The SSM independent analysis is performed in section 2.
In section 3 we discuss its implications for new solar experiments.
Conclusions are drawn in section~4.
In appendix~A we describe the details of the computation.
In appendix B we discuss how KamLand and neutrino factories can test a
high value of $\Delta m^2_{12}\circa{>}10^{-4}\eV^2$,
allowed by solar data in presence of
an undetected systematic error in the Chlorine experiment.

\medskip

Fitting the
solar, atmospheric and LSND anomalies with neutrino oscillations
consistently with all bounds would
require more than 3 neutrinos and peculiar models.
We limit ourselves to oscillations between the 3 SM active neutrinos and we
await for
a confirmation of the LSND result~\cite{LSND}, disregarded in the following.
We use the same notations as in~\cite{lungo}.
The three neutrino masses $m_i$ are ordered such that $\Delta
m^2_{23}>\Delta m^2_{12}>0$ where
$\Delta m^2_{ij}\equiv m_j^2-m_i^2$.
The neutrino mixing matrix is parameterized as
\begin{equation}\label{eq:Vparam}
V =
R_{23}(\theta_{23})\diag(1,e^{i\phi},1)
R_{13}(\theta_{13})
R_{12}(\theta_{12})
\end{equation}
where $R_{ij}(\theta_{ij})$ represents a rotation by $\theta_{ij}$ in the
$ij$ plane,
 $0\le \theta_{ij}\le \pi$,
and $\phi$ is a CP-violating phase.
With these notations, $\theta_{23}$ and $\Delta m^2_{23}\approx \Delta
m^2_{13}$
are relevant to the atmospheric neutrino anomaly,
$\theta_{12}$ and $\Delta m^2_{12}$ to the solar anomaly, while $\theta_{13}$
can affect both solar and atmospheric neutrinos.

\begin{figure}[t]
\begin{center}
\includegraphics[width=8cm]{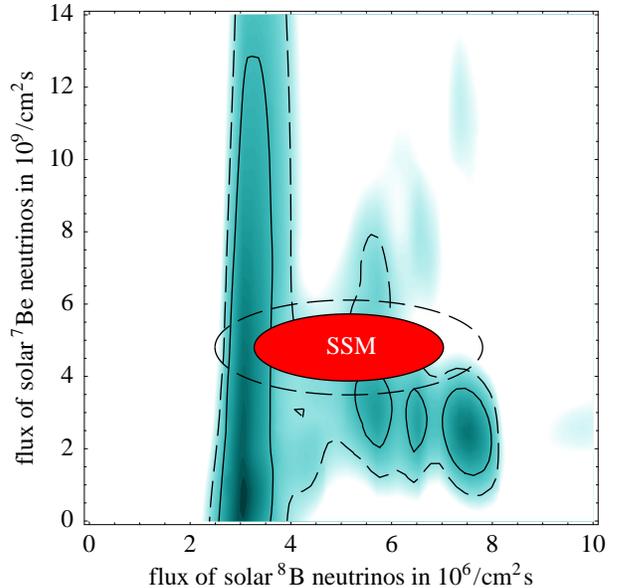}
\caption[SP]{\em Best fit values of the neutrino solar fluxes, as obtained
by this analysis,
 compared with SSM theoretical predictions.\label{fig:SSMindep}}
\end{center}\end{figure}

\begin{figure*}[t]
\begin{center}
\includegraphics[width=18cm]{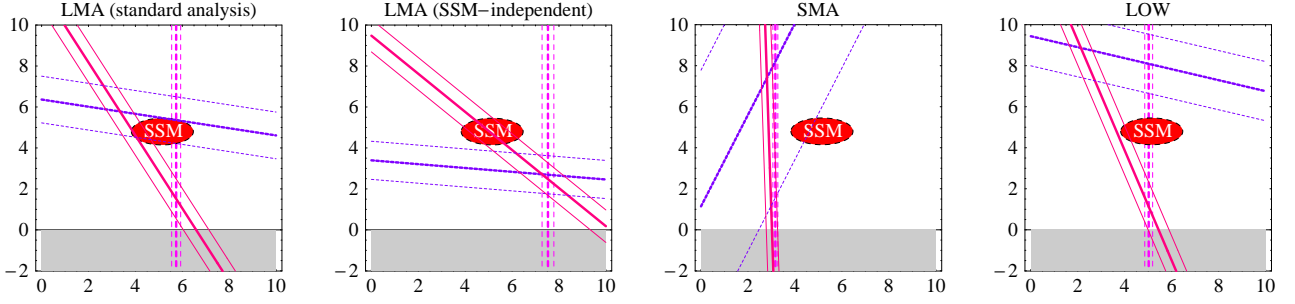}
\caption[SP]{\em Values of the solar neutrino fluxes $(\Phi_{^8\rm
B},\Phi_{^7\rm Be})$ measured by
the Chlorine experiment (continuous lines), the Gallium experiment (dashed
lines)
and by the SuperKamiokande experiment (long dashed lines) assuming four
neutrino oscillation schemes:
$\bullet$ the standard best-fit LMA point in fig.~\ref{fig:incroci}a;
$\bullet$ the solar-model independent best-fit LMA point in
fig.~\ref{fig:incroci}b;
$\bullet$ the best-fit SMA point in fig.~\ref{fig:incroci}c;
$\bullet$ the best-fit LOW point in fig.~\ref{fig:incroci}d.
All four plots have the flux of $^8\rm B$ neutrinos in $10^6 \cm^{-2}{\rm
s}^{-1}$ on the horizontal axis
and the flux of $^7\rm Be$ neutrinos in $10^9 \cm^{-2}{\rm s}^{-1}$ on the
vertical axis.
The ellipse is the  Standard Solar Model prediction.
All errors correspond to one standard deviation.
\label{fig:incroci}}
\end{center}\end{figure*}

\section{SSM independent analysis}
One can perform a {\em useful} almost SSM independent
analysis~\cite{SSManalyses,lungo}
by just treating the overall $^8\rm B$ and $^7 \rm Be$ fluxes as unknown
quantities, to be extracted from the data.
Here we briefly recall how this procedure is justified (see~\cite{lungo}
for explanations and references).
First, it is safe to use the standard spectral functions for the energy
distributions of the single components
$$
i=\{\rm pp, ~~p\hbox{$e$}p,
~~^7\!Be,~~^{13}\!N,~~^{15}\!O,~~{}^{17}\!F,~~{}^8\!B,~~ h\hbox{$e$}p\}$$
of the solar neutrino flux, while the total flux $\Phi_i$ of each component
is regarded as unknown.
Second, it is safe to set to their standard values the ratios
$\Phi_{^{13}\!\rm N}/\!\Phi_{^{15}\!\rm O}$ and
$\Phi_{{\rm p}e{\rm p}}/\!\Phi_{\rm pp}$,
to neglect $^{17}$F neutrinos and
to consider hep neutrinos only when computing the upper tail of the energy
spectrum of recoil electrons in SK.
Although to a somewhat lesser extent, it is also safe to set
$\Phi_{^{13}\!\rm N}/\!\Phi_{^{7}\!\rm Be}$ to its standard value.
Finally the solar luminosity
constraint allows to express the pp flux in terms of the remaining
free parameters $\Phi_{^{7}\!\rm Be}$
and $\Phi_{^{8}\!\rm B}$.

\begin{figure*}[t]
\begin{center}
\begin{picture}(16,1)
\put(3.5,0){\fig{fit}a: standard fit}
\put(10,0){\fig{fit}b: solar-model-independent fit}
\end{picture}
\includegraphics[width=16cm]{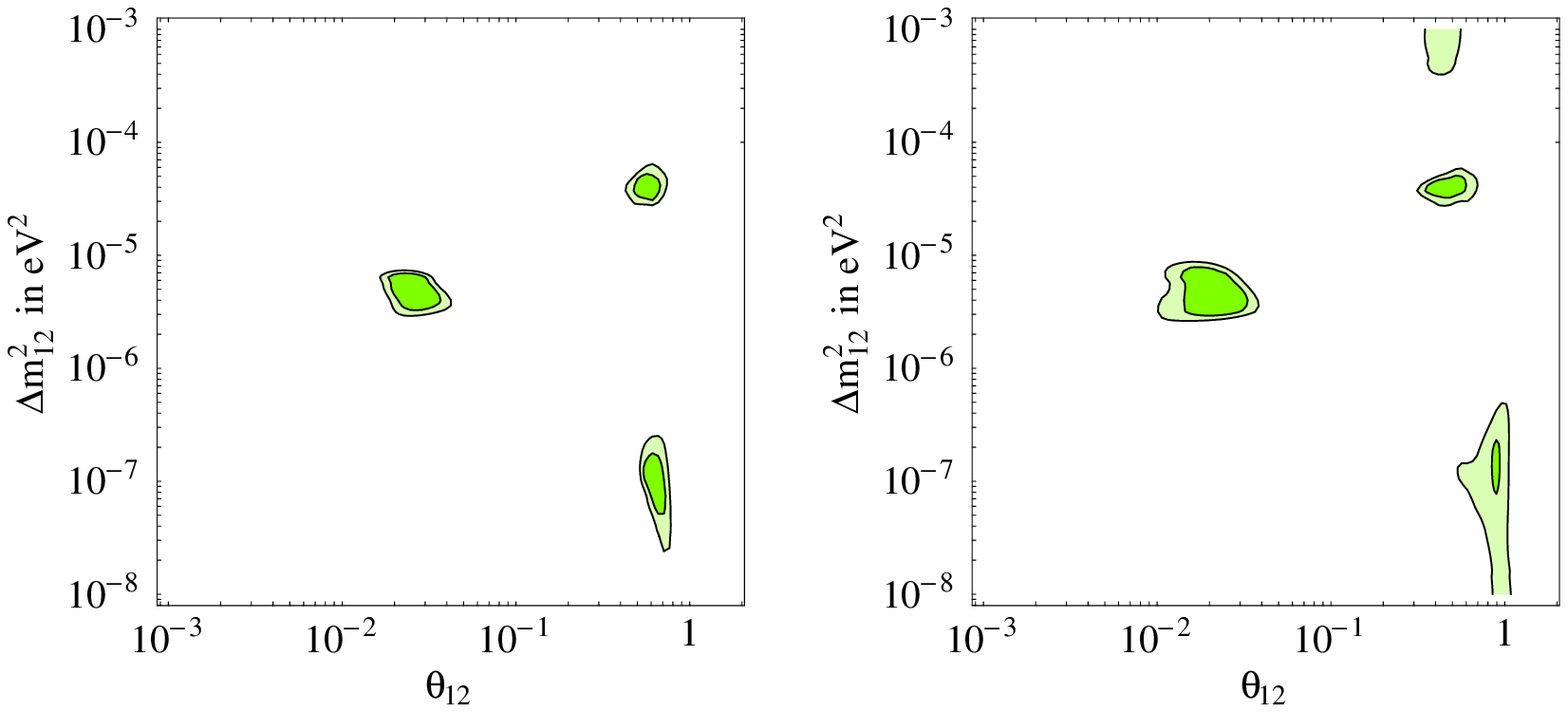}
\caption[SP]{\em Best fit value of the oscillation parameters (a) standard fit
(b) solar-model-independent fit.\label{fig:fit}}
\end{center}\end{figure*}

\medskip

This kind of analysis is useful because solar neutrino rates have been
measured by {\em three} different kinds of experiments.
For any given oscillation pattern, each measured rate gives an allowed band
in the ($\Phi_{^8\rm B},\Phi_{^7\rm Be})$ plane
(few examples are shown in fig.\fig{incroci}).
Requiring a crossing of  all the three experimental bands selects specific
oscillation patterns.
In this way one converts experimental data into informations on the
oscillation parameters
{\em and} on the neutrino fluxes $\Phi_{^8\rm  B}$ and $\Phi_{^7\rm
Be}$. This kind of analysis will become more powerful when the SNO and
Borexino experiments will present their data.
Already now the results are much more restrictive than two years
ago~\cite{lungo}.
SK and Gallium experiments have measured more precisely their fluxes and
the new SK data
now exclude in a SSM independent way a large part of the oscillation
parameter space where MSW effects are large.

Furthermore, the CHOOZ bound on $\bar{\nu}_e$ disappearance~\cite{CHOOZ}
now holds for
all values of the mass splitting  $\Delta m^2_{13}$ allowed by the SK
atmospheric data.
This was not the case one year ago, and implies that $\theta_{13}$ is
small, $\theta_{13}< 15^\circ$ at 95\% C.L.
Therefore $\theta_{13}$ can only have a minor impact on solar neutrino
experiments.
Unless otherwise indicated we will assume that $\theta_{13}=0$.

\medskip

The best-fit values of the neutrino fluxes $\Phi_{^8\rm  B}$ and
$\Phi_{^7\rm Be}$ are shown in fig.\fig{SSMindep}.
The regions delimited by continuous (dashed) lines give the best fit at
90\% (99\%) C.L.\footnote{The
contour lines are drawn at $\Delta \chi^2$ levels that correspond to $90\%$ and
$99\%$ C.L., if one converts values of $\Delta \chi^2$ into ``best fit
probabilities'' $p$ using
the standard expressions valid for a gaussian probability distribution.
This is  not a good approximation since,
as frequently happens in solar neutrino fits, one finds few separate
best-fit solutions,
while a gaussian would have only one peak.
A proper treatment would shift the values of $1-p$ by relative  ${\cal
O}(1)$ factors.
A comparable shift would arise if we performed an exact marginalization of
the joint probability distribution with respect to the oscillation
parameters.
We neglect such corrections, since they are comparable to the uncertainties of Bayesian
inference arising from the need of choosing some prior distribution function.}
The ellipses represent the $90\%$ and $99\%$ C.L.\ SSM prediction for these
fluxes~\cite{BP98}:
\begin{eqnsystem}{sys:SSM}
\Phi_{^8\rm  B}|_{\rm SSM} &=& 5.15~ (1^{+0.19}_{-0.14})\cdot
10^{6}/\cm^2{\rm s},\\
\Phi_{^7\rm Be}|_{\rm SSM} &=& 4.8 (1\pm 0.09)\cdot 10^{9}/\cm^2{\rm s}
\end{eqnsystem}
A standard analysis would include these theoretical constraints in the
$\chi^2$, forcing
$\Phi_{^8\rm  B}$ and $\Phi_{^7\rm Be}$ to be close to the SSM predictions.

\smallskip

Fig.\fig{SSMindep} shows that the best fit regions are neither far from the
SSM predictions of eq.~(\ref{sys:SSM})
nor peaked around them.
This reflects the fact that oscillation patterns that gave the best standard
fits of the measured neutrino rates are now disfavoured
by the SK SSM-independent data.
Basically there are two distinct best-fit regions in fig.\fig{SSMindep}:
\begin{itemize}
\item A region with $\Phi_{^8\rm B}>5 \cdot 10^6/\cm^2{\rm s}$ and
$\Phi_{^7\rm Be}<5 \cdot 10^9/\cm^2{\rm s}$
produced by values of the mixing parameters around the LMA solution.
Fig.\fig{incroci}b shows how a perfect crossing of the three experimental
bands occur
 around $\Phi_{^8\rm B}\approx 7.5 \cdot 10^6/\cm^2{\rm s}$ and
$\Phi_{^7\rm Be}\approx 3\cdot 10^9/\cm^2{\rm s}$. This
crossing is obtained for
$\Delta m^2_{12}=4\cdot 10^{-5}\eV^2$ and $\theta_{12}=0.42$.
The standard analysis requires a crossing
 centered around the SSM prediction:
the best fit is obtained for a slightly larger values of
$\theta_{12}$ and gives the worse crossing shown in fig.\fig{incroci}a.

\item A region with $\Phi_{^8\rm B}\in[2.5\ldots 4] \cdot 10^6/\cm^2{\rm
s}$ produced by
values of the mixing parameters around the SMA solution.
The best crossing, obtained for $\theta_{12}=0.025$ and
$\Delta m^2_{12}=0.5\cdot10^{-5}\eV^2$ is shown in fig.\fig{incroci}c.
It also gives the best standard fit.
The previous best standard fit had larger $\theta_{12}=0.04$ and gave a
crossing
perfectly centered on
the SSM prediction (see fig.~1c in~\cite{lungo}),
but is incompatible with
the day/night and spectral SK data.

\end{itemize}
Oscillation patterns around the LOW region (i.e.\ the one with large
$\theta_{12}$
and $\Delta m^2\circa{<}10^{-7}\eV^2$) give a modest fit
 with $\Phi_{^8\rm B}\approx (4\div 5) \cdot 10^6/\cm^2{\rm s}$ (see
fig.\fig{incroci}d).

\medskip

Before going on, it is useful to consider
 the region around $\Phi_{^8\rm B}\approx 3 \cdot 10^6/\cm^2{\rm s}$ and
$\Phi_{^7\rm Be}\approx 0$.
This region appears due to a unfortunate weakness of our SSM-independent
analysis:
assuming no oscillations, the three bands perfectly cross at
$\Phi_{^8\rm B}\approx 3 \cdot 10^6/\cm^2{\rm s}$ and $\Phi_{^7\rm Be}$
slightly negative.
Therefore the no-oscillation case cannot be excluded at a high confidence
level and
various oscillations patterns not much different from the no-oscillation
case provide acceptable fits.
We consider such crossings as unfortunate accidents.
Before fitting the mixing angles, we exclude by hand such cases by imposing
$\Phi_{^8\rm B} > 0$ and $\Phi_{^7\rm Be}> 1.5~10^9/{\rm cm}^2\,{\rm s}$
rather than $\Phi_{^7\rm Be}> 0$.
This does not conflict much with our purpose of performing a
SSM-independent analysis,
since very low values of the $^7\rm Be$ flux are unphysical, as
the Boron neutrinos, seen in SK, originate from the Berillium ones to a
large extent.

\setlength{\unitlength}{1cm}
\begin{figure*}[t]
\begin{center}\begin{picture}(16,5.5)
\put(0,0){\includegraphics[height=5.5cm]{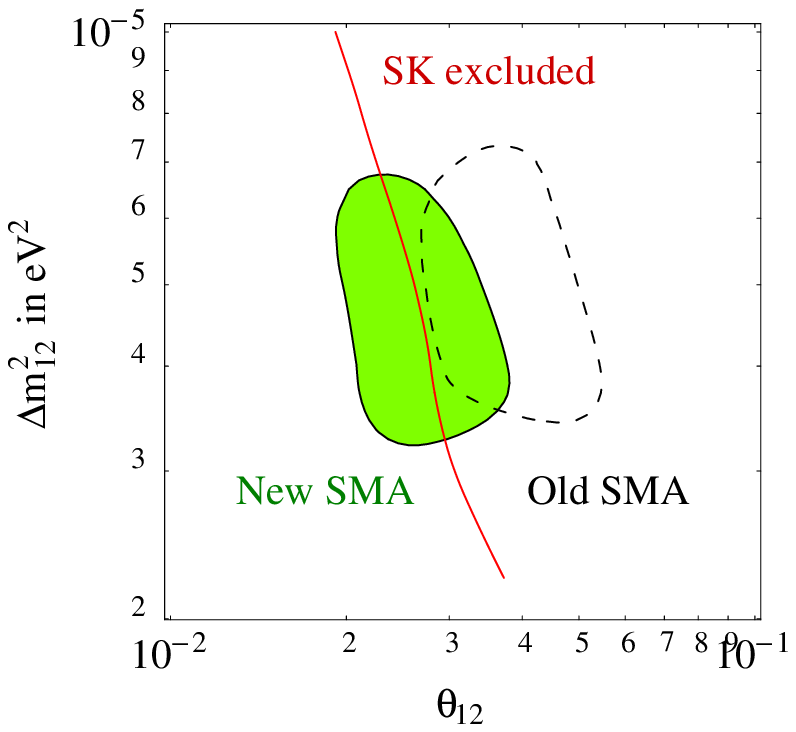}
\includegraphics[height=5.5cm]{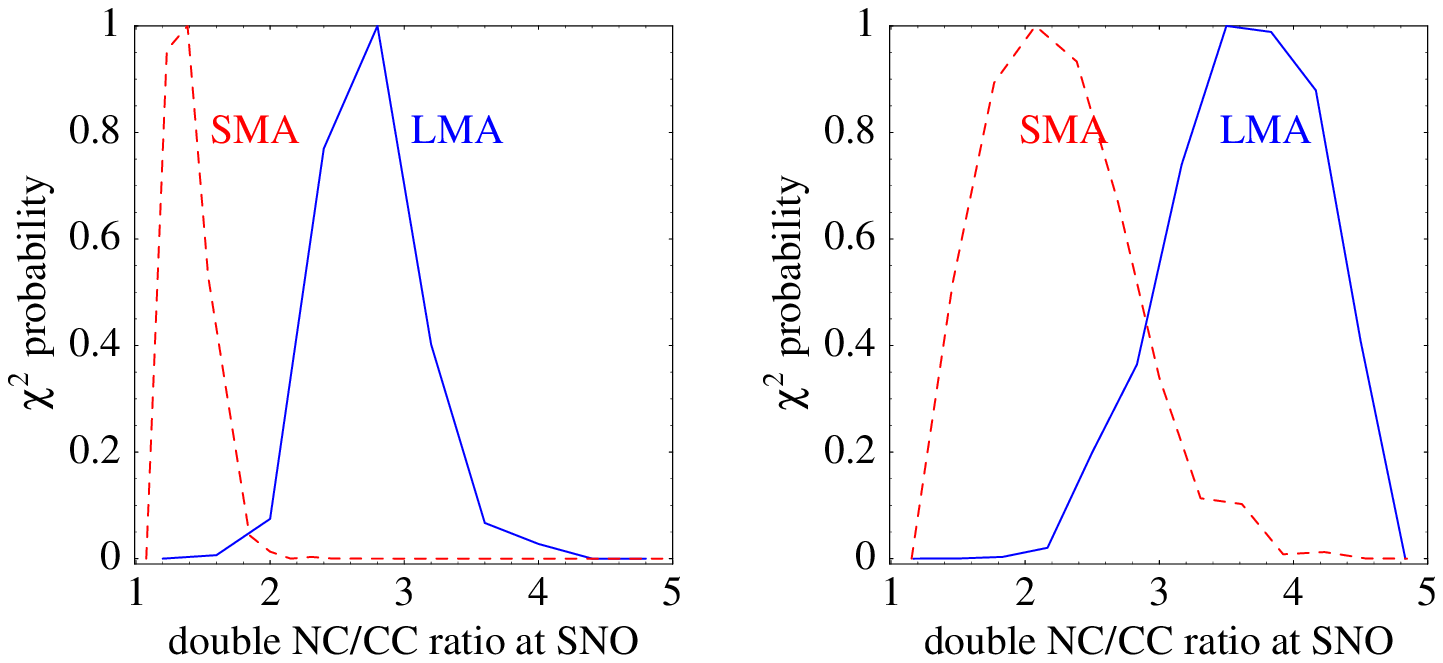}}
\put(1,5.5){\fig{NC/CC}a: old SMA vs new SMA}
\put(7.8,5.5){\fig{NC/CC}b: new fit}
\put(13.8,5.5){\fig{NC/CC}c: old fit}
\end{picture}
\caption[SP]{\em Fig\fig{NC/CC}a:
how the SK bounds on spectral and day/night effects (continuous line) shift
the best-fit SMA region:
the dashed area shows the old SMA, the gray area the new SMA.
All lines correspond to best fit-bounds at $90\%$ C.L.
Fig.s\fig{NC/CC}b,c show the probability distributions (see text) for the
double NC/CC ratio
in the new (fig.\fig{NC/CC}b) and old  (fig.\fig{NC/CC}c) SMA and LMA best
fit regions.
\label{fig:NC/CC}}
\end{center}\end{figure*}

\medskip

Fig.\fig{fit} shows the fit in the usual plane of
 the mixing parameters $\theta_{12}$ and $\Delta m^2_{12}$:
the standard fit is shown in fig.\fig{fit}a and the
the solar-model independent fit in fig.\fig{fit}b.
The best fit regions are not entirely restricted to
$\theta_{12}<\pi/4$~\cite{dark}.

Concerning the standard fit, we mention an important detail not immediately
apparent from the figure.
Like other standard fits~\cite{recentfits}, our fit still contains a SMA
region,
even if the SK spectrum and day/night data have excluded
the `old SMA' region with larger $\theta_{12}$.
The `old SMA' gave such a good standard fit of solar rates that
values of $\theta_{12}$ previously regarded as `too small' now give an
acceptable standard fit of solar rates.
This is why we obtain a new best-fit SMA region, more SMA than the old one
(see fig.\fig{NC/CC}a).
Fig.\fig{incroci}c explains why
such smaller values of $\theta_{12}$ were discarded in old standard fits
but not in old SSM-independent fits:  they give a good crossing of the
three experimental bands,
but at a value of the Boron flux smaller than the one predicted by the SSM.
This means that the SK spectrum and day/night data are not a
problem for the solar-model independent SMA region.\footnote{This
discussion implies
that the sterile neutrino interpretation of the solar neutrino anomaly is
disfavoured by SK only if one imposes the SSM value of the Boron flux.}.
Non-zero values of $\theta_{13}$ just below the CHOOZ bound
slightly shift the crossing point towards higher Boron fluxes,
and therefore slightly improve the quality of the standard fit in the SMA
region.

\smallskip

Presently the LOW solution  gives a better standard fit than the
SMA solution~\cite{recentfits}.
Fig.\fig{incroci}d shows how the three experimental bands cross in the case
of the `best standard fit' LOW solution,
$\theta_{12}=0.66$ and $\Delta m^2_{12}=0.8~10^{-7}\eV^2$.
The crossing is not good, but roughly centered on the SSM prediction.
A solar-model independent analysis does not reward this property.
The best SSM-independent fit in the LOW region has larger $\theta_{12}$ and
lower $\Phi_{^7{\rm Be}}$ than in the standard fit.

The band corresponding to the Ga experiment in fig.\fig{incroci}d
(the almost horizontal one) is not unacceptably high because
Earth-regeneration
effects strongly affect neutrinos with
energies  $E_\nu \approx \MeV (\Delta m^2/4~10^{-7}\eV^2)$.
Unfortunately radiochemical experiments, which detect such neutrinos,
cannot study day/night effects.
Earth-regeneration gives a $\circa{<} 10\%$ seasonal variation of the capture
rate in GNO, since at Gran Sasso nights are longer in winter than in
summer~\cite{GNOseasonal}.
Gallex does not see such an effect.
Present data from all Gallium experiments could be sensitive to a $10\%$
seasonal variation.

\section{Expectations for SNO}
The fact that the SMA solution has migrated toward smaller values of
$\theta_{12}$,
previously considered only in SSM-independent analyses~\cite{lungo},
has significant implications for the SNO experiment.
Previous studies of the significance of the SNO
experiment~\cite{SNOobservables}
found that only a global fit of various
SNO precision observables could eventually discriminate between the SMA and
the LMA solutions.

On the contrary, we think that it is quite possible at SNO to discriminate
between the new LMA and the new SMA solution\footnote{This observation was also
made in a recent
paper~\cite{lastminute}.} or even find evidence for oscillation patterns
suggested by our solar-model
independent analysis by making use
of the NC/CC double ratio
$$r \equiv \frac{\rm (NC~rate)/(CC~rate)}{\rm (NC~rate)/(CC~rate)_{\rm no~oscillation}}.$$
$r$ is an interesting observable because dominant theoretical uncertainties
cancel out when taking the double ratio,
and because expected oscillation effects can be very large
so that one does not need a very precise determination.

In order to perform a quantitative analysis, we compute
the values of $r$ and of the $\chi^2$-probability $$p \equiv
\exp(-\Delta\chi^2/2)$$
for a grid of oscillation patterns in the SMA and in the LMA regions.
The $\Delta\chi^2$ is computed with respect to the local LMA or SMA minimal
$\chi^2$,
so that $p=1$ in the best-fit SMA point and in the best-fit LMA point.
We assume an energy threshold $T_e > 5 \MeV$ on the recoil electrons
originating from CC interactions $\nu_e{\rm d}\to {\rm pp}e$,
but our final results do not depend on this choice.

In fig.\fig{NC/CC}b,c we plot $p(r)$, the maximal value of $p$ at which any
value of $r$ can be
reached in the LMA and in the SMA region.
The $\chi^2$ is computed performing a standard analysis.
In fig.\fig{NC/CC}b we have included the most recent SK data, while in
fig.\fig{NC/CC}c we have not included them.
Fig.\fig{NC/CC}b shows that there is now a neat separation between SMA and
LMA predictions:
measuring $r<2$ or $r>2$ would have clear implications.
A measurement of $r$ can also provide a signal for non standard solutions.
Dividing the possible values of $r$ in 5 distinct ranges, we can summarize
the situation in the following way:
\begin{itemize}
\item[1.] values of $r$ very close to 1  are allowed in the non standard
part of the SSM-independent SMA region
(with smaller $\theta_{12}$ and low $^8{\rm B}$ flux).
Oscillations into sterile neutrinos would also give $r=1$.
\item[2.] $1<r <2 $ is allowed in the standard or SSM-independent SMA region.
\item[3.]  $r$ very close to 2 is the value predicted by the non standard
solution with
high $\Delta m^2_{12}$ and $\theta_{12}\approx
\pi/4$~\cite{lungo,HighDeltam^2}, allowed in presence of
an undetected systematic error in the Chlorine experiment.

\item[4.] $2<r<4$ is allowed in the standard or SSM-independent LMA region;
\item[5.] $4 < r < 5$ is still allowed in the non standard part of the
LMA region;

\end{itemize}
For completeness it should be said that values of  $r$ between 1.5 and 3 are
expected also
in the standard LOW region.

The SNO experiment will improve also  the experimental knowledge of the
solar neutrino fluxes.
The NC rate is not affected by oscillations between active neutrinos,
and therefore provides a measurement of the Boron flux.
It is expected to have a $\circa{<}10\%$ systematic error,
mainly due to the uncertainty in the detection cross
section~\cite{SNOobservables}.
Due to the large spread between the values of the Boron flux required by
the different oscillation patterns (see fig.\fig{SSMindep})
this measurement will also have a significant impact.
In particular the SMA solution requires low values of the Boron flux.

\medskip

The Borexino experiment will be mainly sensitive to the Berillium component
of the solar neutrino flux.
Therefore its data will be represented by one quasi-horizontal band in
fig.s\fig{incroci}, at a level dependent
on the actual oscillation pattern.
Although the main features can be seen already from fig.s\fig{SSMindep}
and\fig{incroci},
a true understanding  of the impact of Borexino data on a SSM-independent
analysis
will require a combined fit of the oscillation parameters and of the
neutrino fluxes.

The KamLand reactor experiment~\cite{KamLand} will measure accurately the oscillation parameters,
if they lie in the LMA region
(see fig.\fig{KamLand}). In this case, the solar neutrino data will give the $^8{\rm B}$ and $^7{\rm Be}$ solar fluxes.
In particular,
the Borexino data will be crucial for an accurate determination of the $^7{\rm Be}$ flux.

\section{Conclusions}

The SK measurements of the energy spectrum and of day/night and seasonal
variations of the
neutrino flux have not realized, so far, any ``smoking gun'' in the study of
the solar neutrino problem.
Nevertheless these measurements provide significant information, since
independent from
theoretical models. Their use, combined with a proper treatment of all the
different rate measurements
allows an almost direct determination of the preferred values of the
$^8{\rm B}$ and $^7{\rm Be}$
solar neutrino fluxes. In turn these values can be compared with the SSM
expectations.

This comparison at present is encouraging but far from conclusive, as
illustrated in fig\fig{SSMindep}.
In particular, also in view of fig.\fig{incroci}, it makes it clear how
premature it is
to select
one specific pattern of neutrino oscillations to explain the solar
neutrinos. Nevertheless,
data expected in the near future especially from SNO or from Borexino
can turn this comparison into a convincing proof of solar neutrino oscillations
and can also provide, at the same time, an independent
 validation of the SSM from neutrino physics.

\paragraph{Acknowledgments}
We thank Andrea Romanino for useful discussions.

\appendix

\section{Details of the computation}
The energy spectra for the independent components of the solar neutrino
flux have been obtained from~\cite{BahcallWWW}.
The neutrino production has been averaged for each flux component
over the position in the sun
as predicted
in~\cite{BP98,BahcallWWW}.
This averaging does not give significant corrections.
MSW oscillations inside the sun have been taken into account in the
following way.
The $3\times 3$ density matrix $\rho_S$ for neutrinos
exiting from the sun is computed using  the
Landau-Zener approximation with the level-crossing probability appropriate
for an exponential density profile~\cite{MSW,Parke}.
The density profile has been taken from~\cite{BahcallWWW} and is
quasi-exponential: small corrections to $\rho_S$ have been approximately
included.
Oscillation effects outside the sun are described by the evolution matrix
$U$, so that
at the detection point  $\rho_{E}=U\rho_S U^\dagger$.
In particular, earth regeneration effects have been computed numerically
using the mantle-core approximation for the earth density profile.
We have used the tree level Standard Model expression for the
neutrino/electron cross section at SK.
The CC and NC cross sections at SNO have been taken from~\cite{ExpResolution}.
The experimental resolution at SK and SNO has been included as suggested
in~\cite{ExpResolution}.

The total neutrino  rates measured with the three
kind of experimental techniques are~\cite{ClSun,KaSun,GaSun,ExpsSun}
\begin{eqnsystem}{sys:S}
\Phi_{\rm Cl}|_{\rm exp} &=& (2.56\pm 0.22)\,\,{10^{-36} {\rm
s}^{-1}}\label{eq:clsig}\\
\Phi_{\rm Ga}|_{\rm exp} &=& (74.7 \pm 5)\,\,{10^{-36} {\rm s}^{-1}}
\label{eq:gasig}\\
\Phi_{\rm SK}|_{\rm exp} &=& (2.40\pm0.08)\cdot 10^6 \,{\rm cm}^{-2}{\rm
s}^{-1}\label{eq:sksig}
\end{eqnsystem}
having combined systematic errors in quadrature with statistical errors.
The SuperKamiokande
experimentalists give directly the value of the flux they measure.
The other experiments involve more uncertain neutrino cross sections and
prefer to
give the frequency of events measured per target atom in their detector.

The solar model independent SK data included in the fit are:
the energy spectrum of the recoil electrons (divided in 18 energy bins
between $5.5\MeV$ and $15\MeV$) and
the total flux measured at SK during the day and during five night bins
(defined according to the value of the cosine of the zenith
angle)~\cite{KaSun,ExpsSun}.
The SK collaboration can include in their fit data about the
zenith angle variation of the recoil electron spectrum
and exclude the old SMA at $95\%$ C.L.
With these unpublished data
the standard and the SSM-independent SMA solution
would be less attractive and fig.\fig{SSMindep} would show a more
neat separation into two distrinct regions.

%

\section{Large $\Delta m^2_{12}$ and nu-factories}
The standard interpretation of the solar neutrino anomaly is based on many
experimental and theoretical ingredients.
We have discussed how the SSM predictions can be tested.
There is one other ingredient that could be not solidly founded and that
has a significant impact on the final
result~\cite{lungo,HighDeltam^2}. Only a single experiment, the Homestake
one, has detected neutrinos with the Chlorine technique
(with the other techniques, data come from two water Cerenkov and two
Gallium experiments).
Furthermore Homestake is the only experiment
that observes a rate  different than one half of the  SSM prediction in absence of oscillations,
therefore excluding an energy-independent survival probability
$P_{ee}(E_\nu)\approx 1/2$.
This important conclusion could be the result of an under-estimation of the
systematic error,
that according to the Chlorine collaboration~\cite{ClSun} is equal to the
statistical error.
It would be interesting to perform a direct calibration of the Chlorine
detector~\cite{Lande}.

$P_{ee}(E_\nu)\approx 1/2$ can be obtained with $\theta_{12}\approx \pi/4$
and $\Delta m^2_{12}\circa{>}10^{-4}\eV^2 $.
This oscillation pattern has
no problems with the recent SuperKamiokande data so that,
even in a standard analysis, it is no longer significantly worse that the new best fits.

The KamLand experiment~\cite{KamLand} will test the LMA region
looking at disappearance of $\bar{\nu}_e$ reactor neutrinos.
If $\Delta m^2_{12}$ and $\theta_{12}$ lie inside the LMA region, KamLand
can accurately measure them~\cite{recent2}.
If instead $\Delta m^2_{12}\circa{>}2 \cdot 10^{-4}\eV^2$, $\bar{\nu}_e$
oscillations are averaged so that a measurement of $\Delta m^2_{12}$
needs a good energy resolution,
a precise knowdlege of the un-oscillated spectra, high statistics and low background.
Assuming that these conditions can be satisfied, fig.\fig{KamLand} shows 
the accuracy  at which KamLand can measure few values of $\Delta m^2_{12}$ and $\theta_{12}$
(represented by the dots)
after three years of running (i.e.\ with 2400 events if no oscillation occur).
If $\Delta m^2_{12}$ is too large
statistical fluctuations often lead to discrete ambiguities in its determination.
A reactor experiment with a {\em shorter\/} baseline would not have this problem.

Here, we study the impact of a large $\Delta m^2_{12}$ at a neutrino
factory~\cite{nu-factory}.
Due to the high beam purity, a neutrino factory will allow
to study $\nu_e\to \nu_\mu$ and $\bar{\nu}_e\to\bar{\nu}_\mu$ oscillations
down to small values of the oscillation probability.
Extensive studies~\cite{nu-factory,rom} have determined the optimal
energy and pathlength that give the maximal sensitivity to a small
$\theta_{13}$.
If $\theta_{13}=0$  `solar' oscillations give effects $\propto (\Delta
m^2_{12})^2$ that can be seen at a neutrino factory if  $\Delta m^2_{12}\circa{>}2.5 \cdot 10^{-4}\eV^2$.
With a non-zero $\theta_{13}$ and
even for values of $\Delta m^2_{12}$ inside the LMA region, `solar' effects
$\propto \Delta m^2_{12}$ have a significant impact on
$\theta_{13}$ measurements at a neutrino factory~\cite{rom}.

\begin{figure}[t]
\begin{center}
\includegraphics[height=7cm]{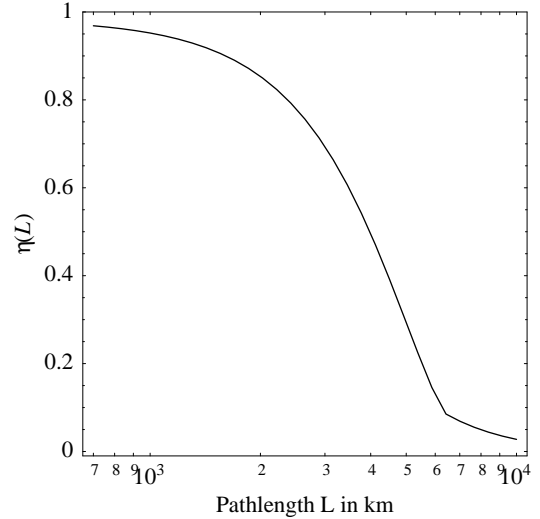}
\caption[SP]{\em Suppression of $\Delta m^2_{12}$ oscillations due to
matter effects as function of the pathlength $L$
\label{fig:eta}}
\end{center}\end{figure}

\begin{figure}[t]
\begin{center}
\includegraphics[height=7cm]{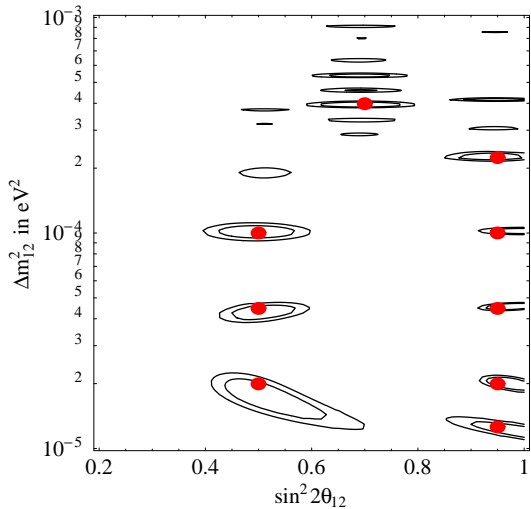}
\caption[SP]{\em Few fits at $90\%$ and $99\%$ C.L. of simulated KamLand data after three years of running.
\label{fig:KamLand}}
\end{center}\end{figure}

The  most promising observable for  discovering $\nu_e\to \nu_\mu$
oscillations is given by
$\mu^-$ appearance  at a relatively short baseline $L\approx 700\,{\rm km}$.
The number $N(\mu_-)$ of $\mu^-$ events produced by both `solar' and
`atmospheric' oscillations
can be approximated by
treating $\theta_{13}$ and $\Delta m^2_{12}$ effects as perturbations. 
This gives\footnotemark[2]\footnotetext[2]{Using the formula $e^{M+\epsilon}=e^M(1+\int_0^1 e^{-xM}\cdot \epsilon\cdot e^{xM}dx + {\cal O}(\epsilon^2)$).}
$P(\nu_e\to\nu_\mu) \approx |\Delta_{e\mu}^{\rm eff} L/2E_\nu|^2$
and 
\begin{equation}\label{eq:shortL}
N(\mu^-) \approx \frac{N_\mu N_{\rm
kt}\epsilon}{10^{21}}\frac{E_\mu}{70\GeV}\left|\frac{\Delta_{e\mu}^{\rm eff}}{10^{-5}\eV^2
}\right|^2
\end{equation}
where $N_\mu$ is the number of $\mu^+$ decays occurring in the straight section
of the storage ring pointing towards the detector,
$E_\mu\sim (20\div 50)\GeV$ is the $\mu^+$ energy,
$N_{\rm kt}$ is the size of the detector in kilo$\,$tons,
$\epsilon$ is the efficiency for the detection of $\mu^-$.
Using the parameterization\eq{Vparam}
\begin{eqnarray*}
\Delta_{e\mu}^{\rm eff} &\equiv & e^{i\phi}\Delta m^2_{12} c_{23} c_{13}  c_{12}s_{12}\frac{e^{iAL/2E}-1}{iAL/2E}+\\
&&+(\Delta m^2_{13}-\Delta m^2_{12}s_{12}^2)s_{13}c_{13}s_{23}\frac{e^{-i A' L/2E}-1}{-iA'L/2E}
\end{eqnarray*}
where $A=2\sqrt{2} N_e G_F E_\nu$, $A'=\Delta m^2_{13}-A$,
$c_{ij}\equiv\cos\theta_{ij}$ and $s_{ij}\equiv\sin\theta_{ij}$.
If the phase factors are large one should average $N(\mu^-)$ over the neutrino spectrum, otherwise one can set $E_\nu\approx E_\mu$.
For small $L$ (in practice for $L\circa{<}700\,{\rm km}$)
$\Delta_{e\mu}^{\rm eff}$ reduces to the $e\mu$ element of the neutrino squared mass matrix.
Assuming a short baseline, $\Delta m^2_{12}\ll\Delta m^2_{13}$, $\theta_{12}\approx\theta_{23}\approx\pi/4$ and $\theta_{13}\ll 1$
$$\Delta_{e\mu}^{\rm eff} \approx \theta_{13}\frac{\Delta
m^2_{13}}{\sqrt{2}}+e^{i\phi}\frac{\Delta m^2_{12}}{2\sqrt{2}}$$
we clearly see in an analytical way how significant $\Delta m^2_{12}$
oscillations can be.
An analogous approximation holds for $\bar{\nu}_e\to\bar{\nu}_\mu$ signals.

At  baselines $L\circa{>}10^3\,{\rm km}$ matter effects become significant.
Using eq.\eq{shortL},
the number of events due to $\Delta m^2_{12}$ oscillations only can be
written as its value at $L=0$ multiplied by the function
$$\eta(L)=\frac{\sin^2 x}{x^2}<1,\qquad x\equiv \frac{N_e G_F L}{\sqrt{2}}$$
which exact numerical value is plotted in fig.\fig{eta}.
Therefore a short baseline $L\approx 700\,{\rm km}$ is optimal for
discovering $\Delta m^2_{12}$ effects.

\medskip

By performing a global fit of simulated nu-factory data we find that it
will be difficult to distinguish  $\Delta m^2_{12}$ effects from
$\theta_{13}$ effects at a good C.L.\
by comparing data taken at different pathlengths and/or different neutrino
energies, as suggested in~\cite{rom}.
For certain values of the CP violating phase
a comparison between $\nu_e\to \nu_\mu$ and $\bar{\nu}_e\to\bar{\nu}_\mu$ rates
allows a better discrimination.
Such `precision studies' are statistically significant only with a
sufficient number of events.
For example, observing few events only would not allow to say if they are
due to a $\theta_{13}\approx 0.007$ around its nominal sensitivity,
or due to a $\Delta m^2_{12}\approx 3~10^{-4}\eV^2$.
In conclusion, if $\Delta m^2 \circa{>} 2~10^{-4}\eV^2$ so that KamLand cannot measure it,
an accurate measurement of $\Delta m^2$ cannot even be obtained with a neutrino factory:
a new reactor experiment with intermediate baseline $\sim 10$ km would be necessary.

If KamLand will give a precise measurement of $\Delta m^2_{12}$
free from discrete ambiguities
(this could not be the case if $\Delta m^2_{12}\circa{>}2 \cdot 10^{-4}\eV^2$),
by combining KamLand data with nu-factory data one can usually obtain a
satisfactory fit  of $\theta_{13}$ and
of the CP violating phase.

\nonfrenchspacing

\newpage

\normalsize

   \begin{figure}[t]
  $$\includegraphics[width=8cm]{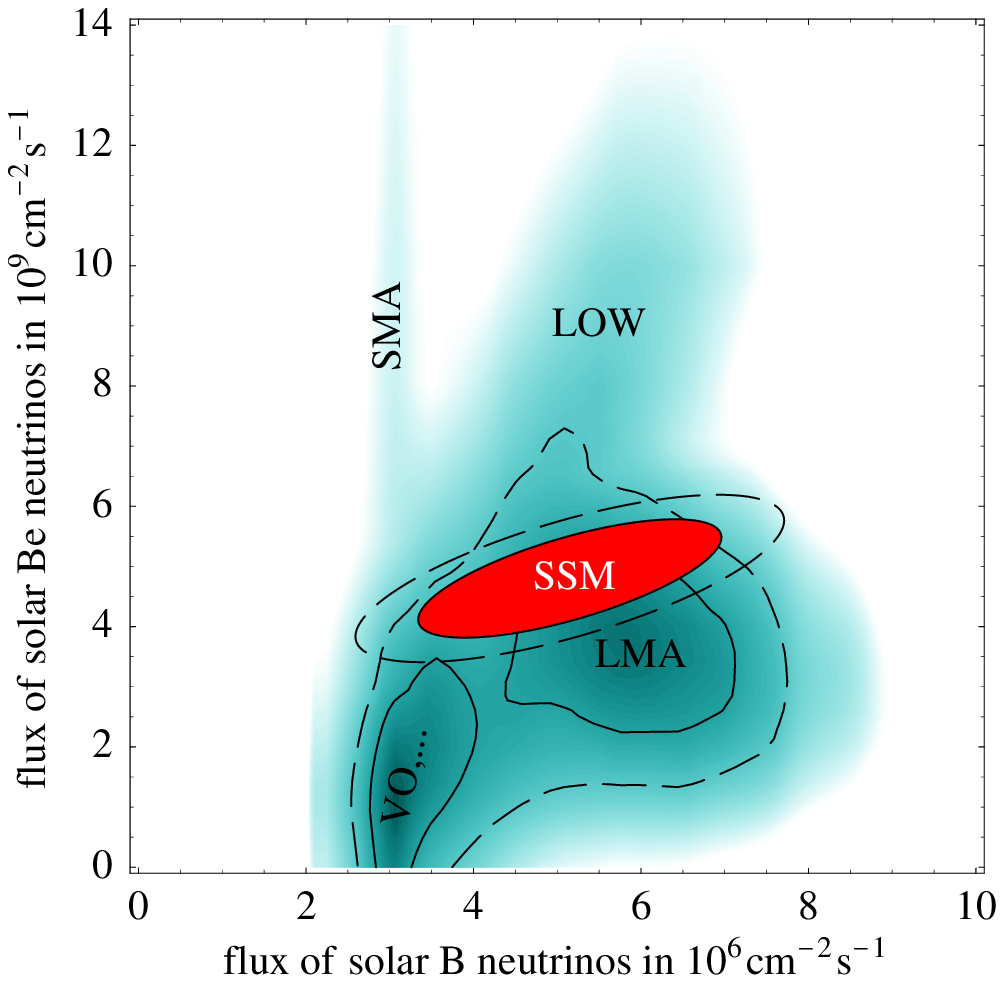}$$
\caption[]{\em Fig.\fig{SSMindep} updated including the SNO CC result:
best fit, at $90\%$ {\rm CL} (continuous line) and $99\%$ {\rm CL} (dashed line), of the neutrino solar fluxes
    compared with SSM theoretical predictions.
\label{fig:todayFlux}}
\end{figure}


\begin{figure*}[t]
$$
\includegraphics[width=55mm]{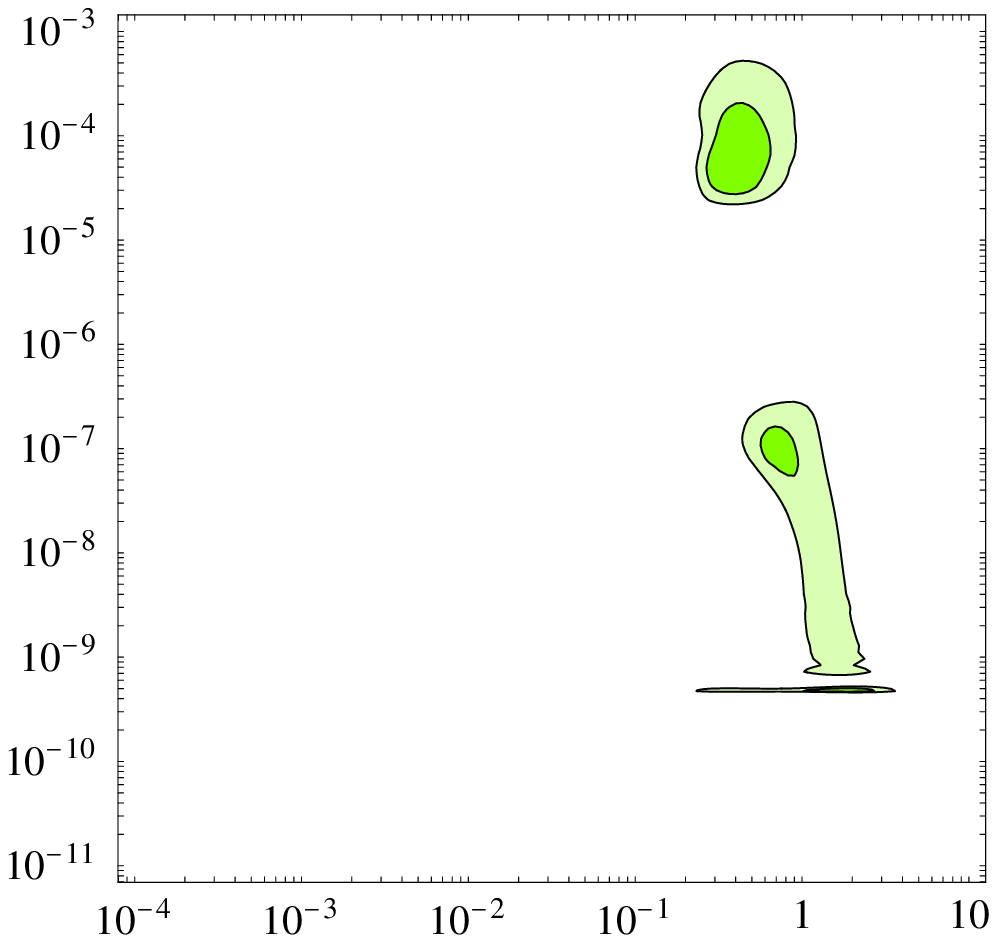}~
\includegraphics[width=55mm]{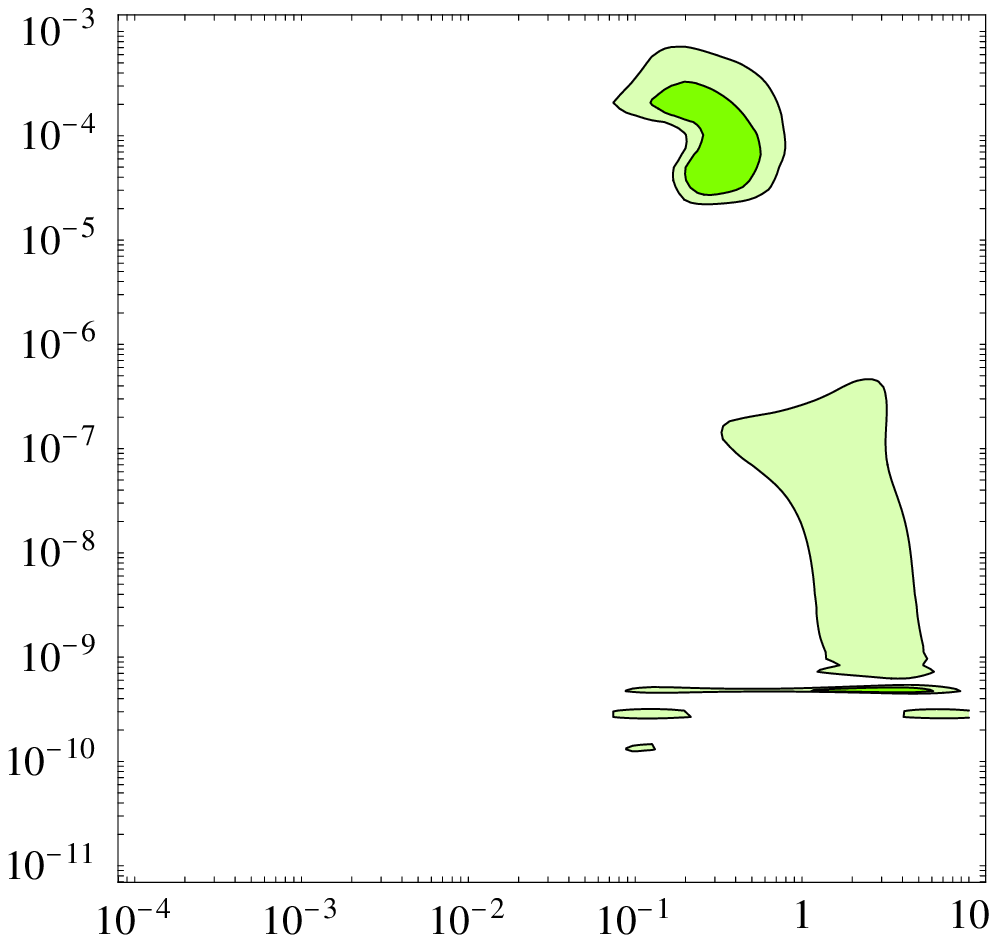}~
\includegraphics[width=55mm]{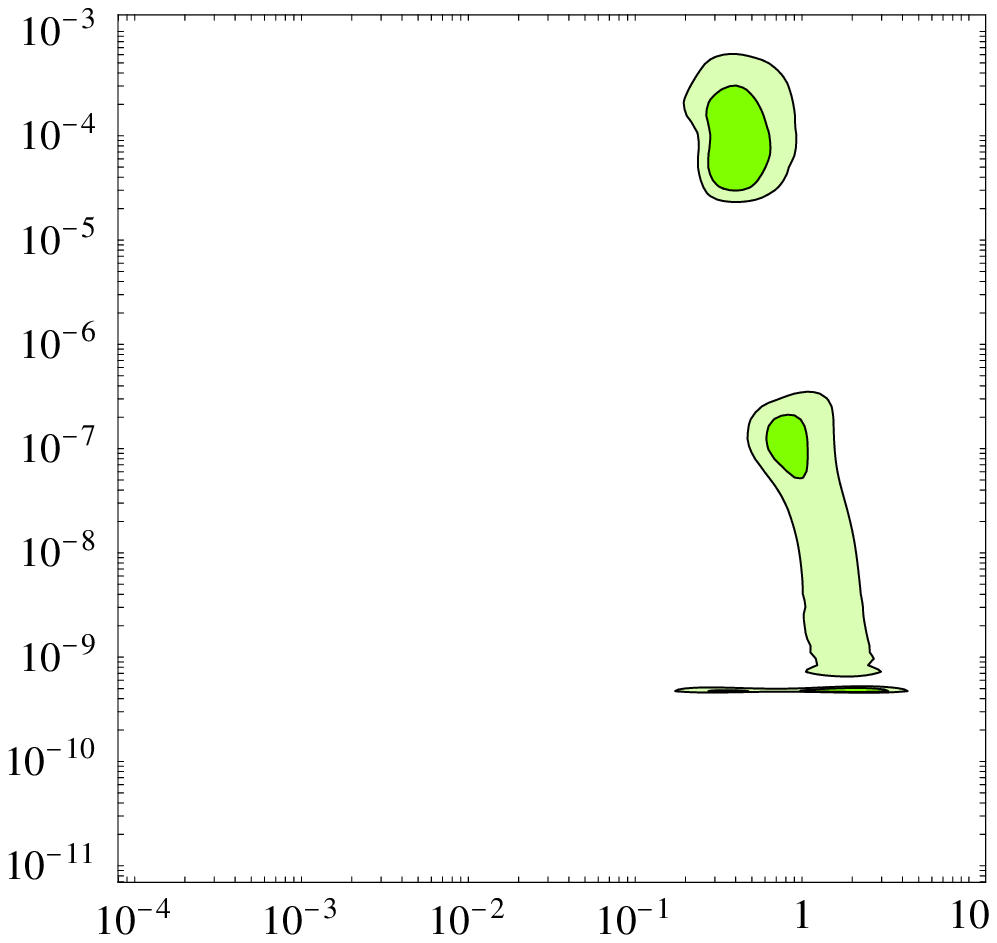}~ $$
  \caption[]{\em Global fits of solar data (updated including SNO CC data) at $90$, $99\%$ {\rm
CL} in the
$(\tan^2\theta_{12},\Delta m^2_{12}/\eV^2)$ plane: (a) standard fit; 
(b) solar model independent fit;
(c) standard fit dropping the uncalibrated Chlorine data.\label{fig:today}}
$$
\includegraphics[width=55mm]{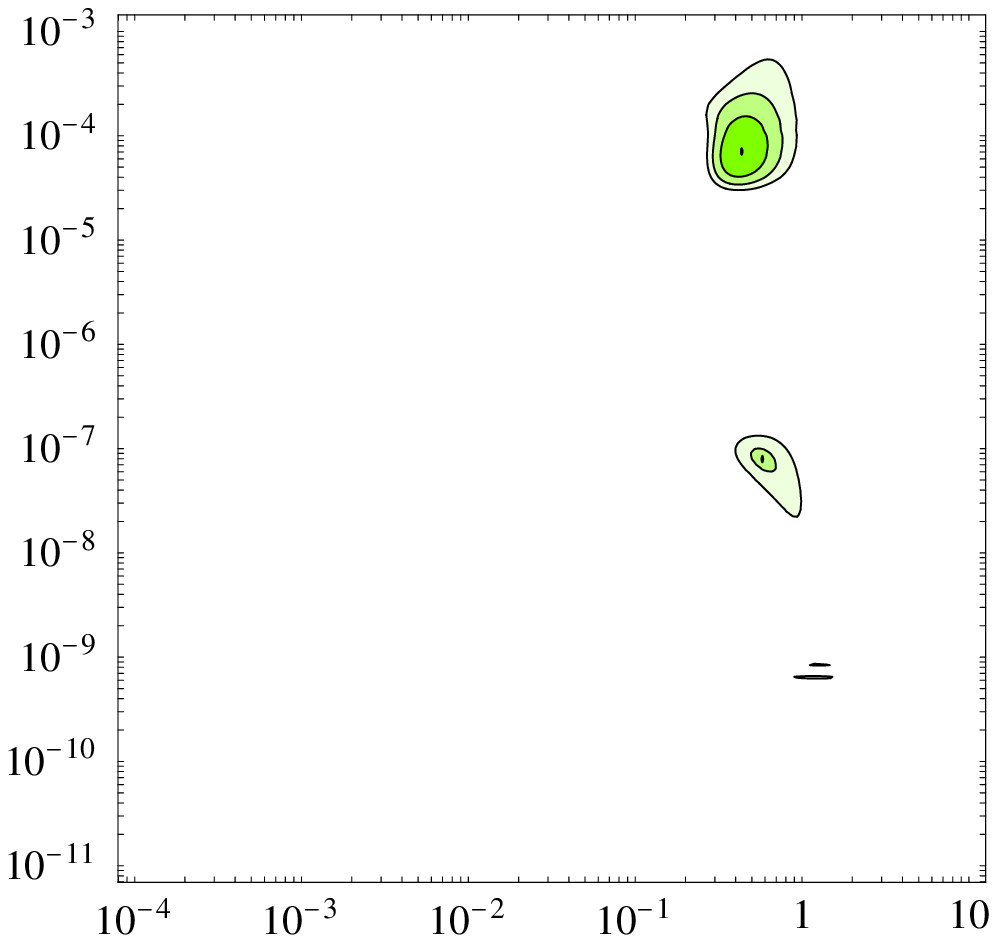}~
\includegraphics[width=55mm]{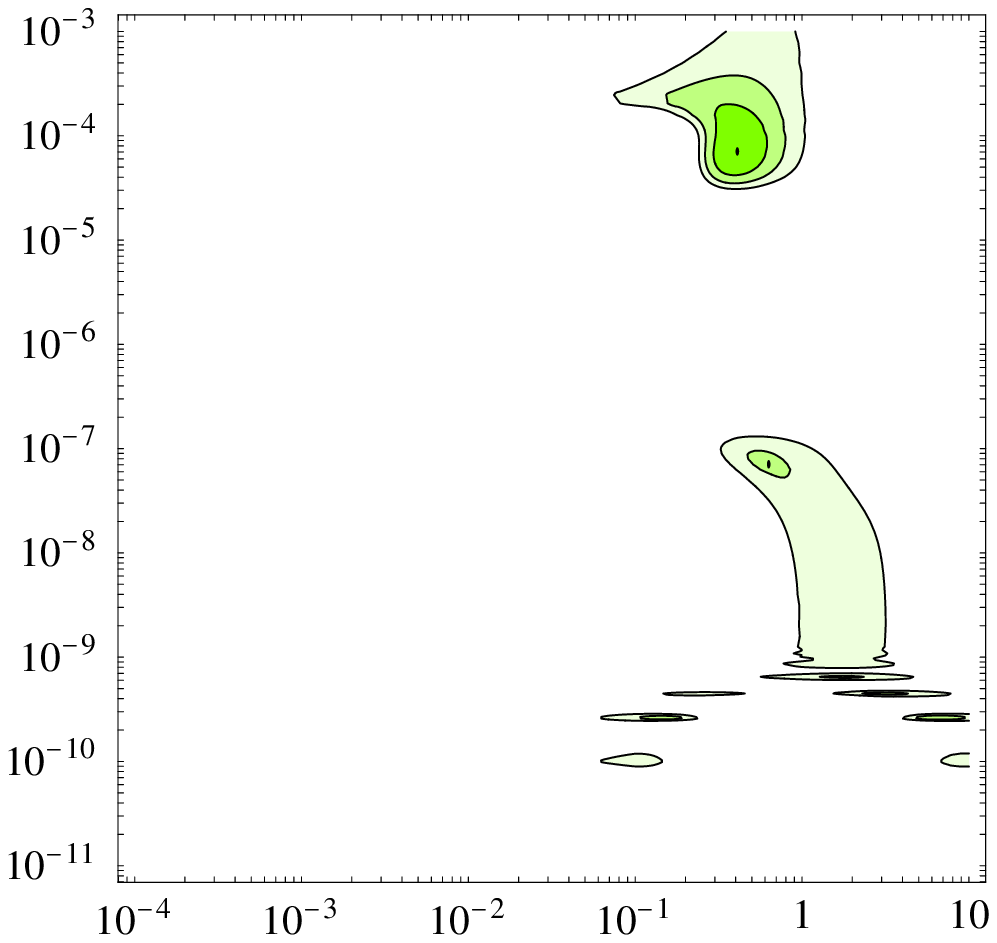}~
\includegraphics[width=55mm]{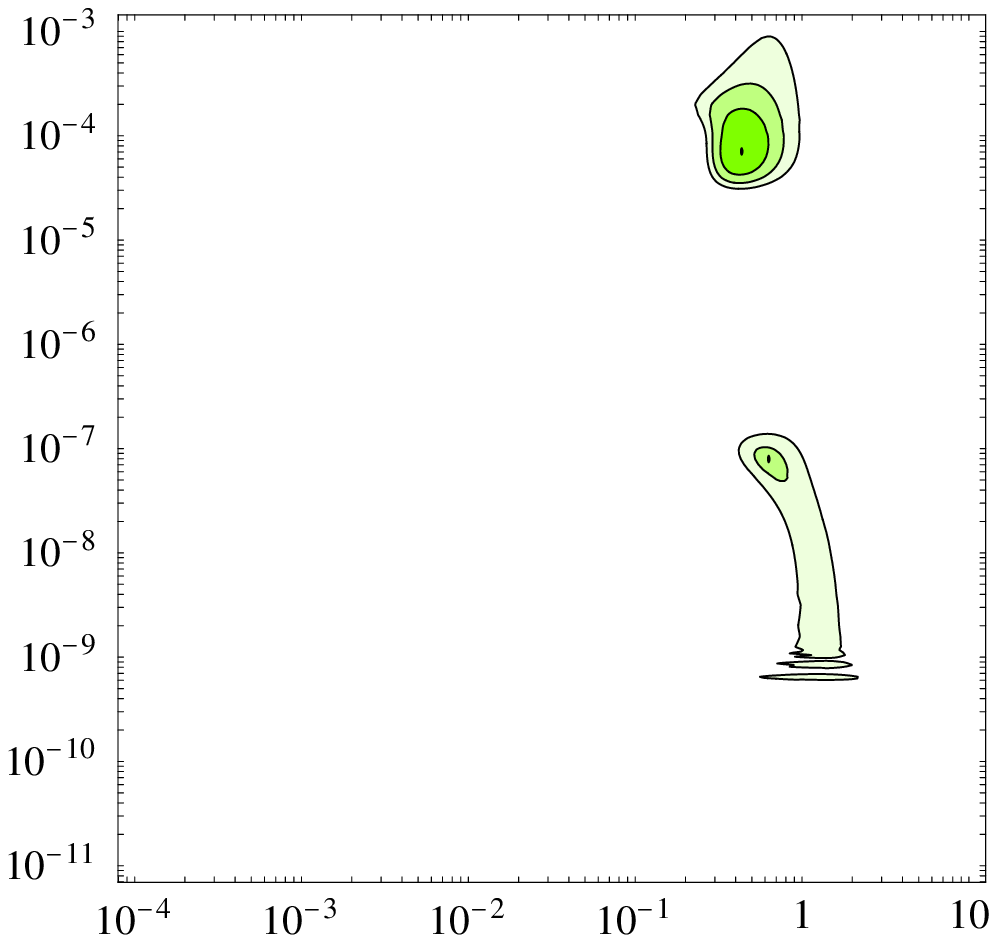}~ $$
  \caption[]{\em Global fits of solar data (updated including SNO NC data) at 
$90$, $99$, $99.9\%$ 
{\rm CL} in the
$(\tan^2\theta_{12},\Delta m^2_{12}/\!\eV^2)$ plane: (a) standard fit; 
(b) solar model independent fit;
(c) standard fit dropping the uncalibrated Chlorine data.\label{fig:today2}}

\end{figure*}

\section*{5~~~Addendum: SNO CC results}\label{6in}
In this addendum, we  update our results by adding the first SNO~\cite{SNO} data.
The SK collaboration published the day + night spectral data~\cite{SKnew} as 19 + 19 energy bins, 
so that we can now include this information in the fit.
We also explicitly included the CHOOZ data~\cite{CHOOZ}, that disfavour values of $\Delta m^2_{12}$ above $0.7~10^{-3}\eV^2$ for large mixing
$\theta_{12}\sim \pi/4$. Finally, we improved the numerical accuracy of our computation, and extended it to include vacuum oscillations~\cite{QVO}.

As pointed out in section~3, the measurement of the NC/CC ratio alone was expected to discriminate between LMA and SMA.
The measured value happens
to fall into case 4.\ (cases 3.\ and 5.\ are not significantly disfavoured).
Therefore the SMA region is now strongly disfavoured,
even from a solar-model-independent point of view.
Furthermore, the solar-model-independent LMA region (case 5.) no longer gives a best $\chi^2$
significantly lower than the standard fit.
These results are confirmed by an updated analysis, as shown in fig.~\ref{fig:todayFlux} and \ref{fig:today}.

In fig.~\ref{fig:todayFlux}  we show the updated fit of the solar neutrino fluxes.
Values of the $^8{\rm B}$ flux detectably different from the SSM prediction no longer give good fits.
On the contrary, a discrepancy between the  $^7{\rm Be}$ flux and its SSM prediction could still have significant effects.
The best-fit region at 90\% CL in fig.~\ref{fig:todayFlux} is composed by two disjoint regions.
The one with smaller fluxes is obtained from vacuum oscillations
(but can also be obtained with LMA, LOW, SMA oscillations with a worse CL).
The one with larger fluxes is obtained from LMA oscillations.

In fig.~\ref{fig:today} we show the updated global fits for the oscillation parameters:
fig.~\ref{fig:today}a shows a standard fit, while
fig.~\ref{fig:today}b is the solar model independent fit described in section~2.
In fig.~\ref{fig:today}c we perform a standard fit, but dropping the uncalibrated Chlorine rate from the fit.
The results of fig.s~\ref{fig:today}a,b,c are quite similar:
more or less acceptable fits can be obtained for
a large range of $\Delta m^2$ and
large mixing angle, while the SMA solution is always disfavoured.

Finally, we mention another important aspect of solar model independent analyses.
It is sometimes said that it is useless to measure the Gallium rate
with an error much smaller than the solar model uncertainty.
As explained in section~2 and illustrated in fig.s~\ref{fig:incroci}, this is not true.
Fig.s~\ref{fig:incroci} in fact shows that the amount of information that can be extracted 
in a solar model independent way from the solar rates
is today limited by the accuracy of the Chlorine experiment.
This limitation will disappear when Borexino will measure the $^7{\rm Be}$ flux.
Fig.s~\ref{fig:today} show that 
solar model independent considerations already now give useful informations on the oscillation parameters.
\label{6out}

\nonfrenchspacing

\newpage

 \begin{figure}[t]
  $$\includegraphics[width=8cm]{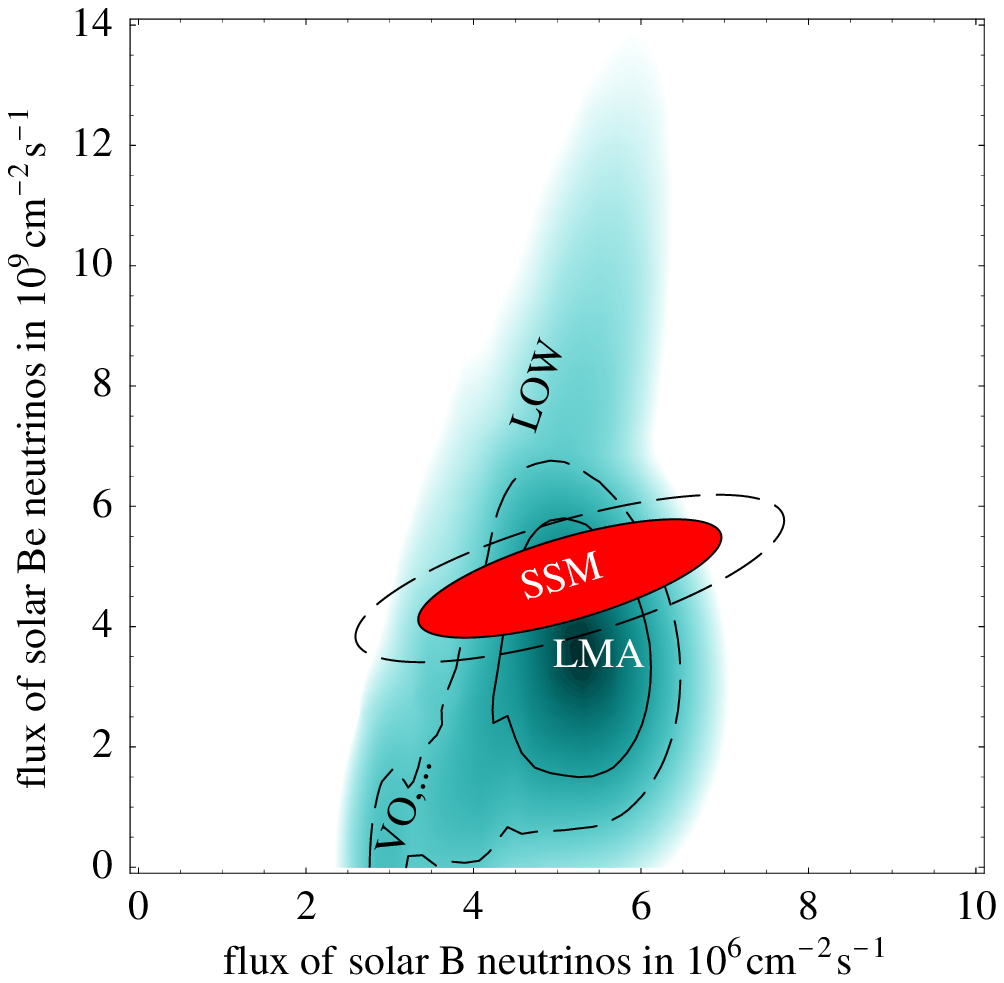}$$
\caption[]{\em Fig.s\fig{SSMindep},\fig{todayFlux}  updated including the SNO NC result:
best fit, at $90\%$ {\rm CL} (continuous line) and $99\%$ {\rm CL} (dashed line), of the neutrino solar fluxes
    compared with SSM theoretical predictions.
\label{fig:last}}
\end{figure}

\section*{6~~~Addendum: SNO NC results}\label{7in}
In this addendum, we  update our results by adding the latest SAGE, GNO and SK data
and the SNO day/night energy spectrum~\cite{newdata}.
As explained in the SNO paper, the SNO spectrum allows to extract its NC, CC and ES contributions.
For example, assuming energy-independent oscillations between active neutrinos,
our reanalysis of SNO data gives
\begin{eqnsystem}{sys:S}
\Phi_{\rm NC} =& \Phi_{^8\rm B}  &= (5.2 \pm 0.5)~10^6/\cm^2{\rm s},\\
\Phi_{\rm CC} =&  P_{ee}\Phi_{^8\rm B} &= (1.76 \pm 0.08)~10^6/\cm^2{\rm s}.
\end{eqnsystem}
(the errors are somewhat anti-correlated).\footnote{Our reanalysis slightly differs from the corresponding SNO analysis
in~\cite{newdata}. SNO extracts from all data (energy and zenith-angle spectra) 
the CC, NC and ES components.
We instead extract the CC, NC components from the energy spectrum,
assuming the standard relation $\Phi_{\rm ES} \approx \Phi_{\rm CC} + 0.15 \Phi_{\rm NC}$
(the ES rate have also been accurately measured by SK).
We take into account systematic errors and backgrounds as described by SNO~\cite{newdata},
as well as the theoretical uncertainty on the Boron energy spectrum
(that also affects other solar neutrino experiments).
See~\cite{pulls} for a useful discussion of these issues.
}

As expected,
new data have a significant impact on the oscillation fit~\cite{AsExpected}.
In fig.~\ref{fig:today2} we show the updated global fits for the oscillation parameters:
fig.~\ref{fig:today2}a shows a standard fit, while
fig.~\ref{fig:today2}b is the solar model independent fit described in section~2.
In fig.~\ref{fig:today2}c we perform a standard fit, but dropping the uncalibrated Chlorine rate from the fit.
The results of fig.s~\ref{fig:today2}a,b,c are quite similar:
more or less acceptable fits can be obtained for
a large range of $\Delta m^2$ and
for large mixing angle, while the SMA solution is always incompatible with data
at about $5\sigma$ level.
In all analyses LMA oscillations give a better fit than
LOW and (Q)VO oscillations.
In fig.s\fig{today2}a,b,c we have not included the CHOOZ bound,
in order to show that $\Delta m^2\circa{>}10^{-3}\eV^2$ is now disfavoured by solar data.

In fig.\fig{last} we show the Boron and Beryllium fluxes,
as extracted from the solar neutrino data
without assuming solar model predictions for these fluxes,
and assuming that the solar anomaly is due to $\nu_e \to \nu_{\mu,\tau}$ oscillations.

Within this framework, SNO has measured the Boron flux through NC scattering,
finding a value in agreement with solar model predictions.
A much larger Boron flux was already excluded in our previous analysis (see figures\fig{SSMindep}
and\fig{todayFlux}) because
it would need a small and energy-independent 
survival probability $P_{ee}(E_\nu)$, that
cannot be obtained by oscillations.

\medskip

In view of this experimental progress,
one can study if new significant information,
e.g.\
about the CNO and $pp$ fluxes, can now be extracted
from a more general analysis.
Since the results are not very interesting,
we just briefly describe their main features.
Only two kind of experiments,
Gallium and Chlorine, have measured low-energy neutrino fluxes
Therefore, one can think of extracting the values of two of these fluxes at most.
The Chlorine experiment is not very sensitive to low energy neutrinos:
after subtracting the $\sim80\%$ Boron contribution, as measured by SNO via CC,
the residual Chlorine rate is just about $2\sigma$ above zero.

%
%
%
%

A first interesting question is: do
solar neutrino data discriminate between CNO and $^7$Be neutrinos?
We find that equally good fits are obtained in the extreme limits of vanishing CNO or $^7$Be flux.
Therefore  fig.\fig{last}, where the Be flux is studied,
provides all the significant information on Be and CNO fluxes,
and implies that their sum is at most 2 times larger than solar model predictions.
This means that the solar luminosity constraint fixes the $pp$ flux to be
around its solar model value.

A second issue is: 
what is it possible to say on the $pp$ flux,
without imposing the luminosity constraint 
(and of course without assuming the solar model predictions for the other fluxes).
The relatively more interesting result is that
the $pp$ flux cannot be zero,
as can be seen by comparing the Gallium, Chlorine and SNO CC rates.

\smallskip

Ref.~\cite{pulls} finds that the pulls between
solar model predictions and their best-fit values
(as obtained from a global oscillation fit that {\em includes} these predictions)
are small.
Note, however, that solar models are not confirmed in this way.
While a large pull would signal a problem,
a small pull is obtained in two cases:
when the experimental determination agrees with the prediction
(this happens in the case of the Boron flux)
but also when the experimental error is much larger than
the theoretical error
(this roughly happens for all other fluxes).
Looking at pulls only, one cannot discriminate these two extreme cases.
In order to test predictions
one must extract the predicted quantities (e.g.\ the B and Be fluxes in fig.\fig{last},
more generically any other `systematics' relevant for solar data)
by fitting the data {\em without} assuming the predictions under examination.

\label{7out}

\end{document}